\documentclass[twoside,twocolumn]{article}
 \pdfoutput=1
\usepackage{fitee}
\usepackage{mathpartir}
\usepackage{listings} 
\usepackage{subfigure}
\usepackage{todonotes}

\graphicspath{ {./} }

\usepackage[colorlinks, breaklinks = true]{hyperref}		
\addto{\captionsenglish}{%
  
}

\begin{document}

\title{A New Hierarchical Software Architecture Towards Safety-Critical Aspects of a Drone System}

\author[$\dagger$$\ddagger$1]{Xiao-rui ZHU}%
\author[1]{Chen LIANG}%
\author[1]{Zhen-guo YIN}%
\author[$\dagger$2]{\authorcr Zhong SHAO}
\author[2]{Meng-qi LIU}%
\author[2]{Hao CHEN}%

\affil[1]{Department of Mechanical and Automation, Harbin Institute of Technology, Shenzhen 518055, China}
\affil[2]{Department of Computer Science, Yale University, New Haven 06511, United States}

\shortauthor{Zhu et al.}	

\authmark{}

\corremailA{xiaoruizhu@hit.edu.cn}
\corremailB{zhong.shao@yale.edu}
\emailmark{$\dagger$}	


\abstract{In this paper, a new hierarchical software architecture is proposed to improve the safety and reliability of a safety-critical drone system from the perspective of its source code.
The proposed architecture uses formal verification methods to ensure that the implementation of each module satisfies its expected design specification, so that it prevents a drone from crashing due to unexpected software failures.
This work builds on top of a formally verified operating system kernel, CertiKOS\citep{gu2015deep}.
Since device drivers are considered the most important parts affecting the safety of the drone system, this paper mainly focuses on verifying bus drivers such as the SPI driver and the I2C driver in a drone system using a rigorous formal verification method\citep{chen2016toward}.
Experiments have been carried out to demonstrate the improvement in reliability in case of device anomalies.}


\doi{10.1631/FITEE.1800636}	
\code{A}
\clc{V279;TP311.5}



\support{Project supported by the National Natural Science Foundation of China (No.91648012) and Shenzhen Science, Technology, and Innovation Comission, China (No. JCYJ20160401100022706)}

\orcid{Xiao-rui ZHU, http://orcid.org/0000-0003-1400-059X Zhong SHAO, http://orcid.org/0000-0001-8184-7649}	
\articleType{}

\maketitle

\section{INTRODUCTION}
In recent years, small unmanned aerial vehicles (UAV) or drones have drawn more and more attention because of their low cost, compact size.
As small UAVs come into our daily life, safety concerns are also rising.
Failures of a drone may result in severe damage to the environment and serious injury to the public\citep{simpson2006safety}.

Aside from maneuver mistakes, software errors in the controller are also one of the main reasons for UAVs failures.
The fault may come from the algorithm itself or its actual implementation\citep{malecha2016towards}.
A lot of work has been done to improve the reliability of UAV systems.
Most efforts have focused on algorithms, such as improving modeling accuracy\citep{gordon2000principles}, enhancing the robustness of control algorithms\citep{lee2010geometric}, and reducing sensor errors\citep{de2012uav}.
In 2013, R{\'e}ti proposed a hardware solution to improve the safety by developing a smart mini actuator which integrated measurements of position and angular rate with controlling microprocessors\citep{reti2013smart}.
Few people so far have addressed bugs in the implementation of algorithms at the source code level.
For a safety-critical real-time system like an UAV, this negligence could result in problems such as loss of synchronization (caused by irregular response from external sensors) and high approximation errors (caused by floating-point computation)\citep{malecha2016towards}.
These problems are subtle but might degrade the performance or even cause the drone to crash.

Formal verification is a technique to conduct correctness proof of a program (or the contradiction if the program contains errors) in accurate and well-formed mathematical and logical constructs.
It is used to prevent subtle errors in the source code of control systems \citep{ricketts2015towards, malecha2016towards, bohrer2018veriphy}.
Preventing such errors would increase the reliability and safety of drone systems.
In 2015, foundational verification techniques in the theorem prover Coq, were applied to a quadrotor system to verify the correctness of two shims (saturation blocks) that were used to limit the velocity and height of the quadrotor\citep{ricketts2015towards}.
In 2016, the same research group verified a runtime monitor in order to provide strong guarantees about maximum velocities and accelerations of a drone\citep{malecha2016towards}.
In 2018, Brandon Bohrer \citep{bohrer2018veriphy} designed a verified pipeline for generating concrete controller code from high-level models. However, these efforts for formally verifying control systems are not enough for a hybrid real-time drone system.

Real-time operating system (RTOS) plays an important role in scheduling real-time processes and interacting with devices.
Traditional RTOSes, including Nuttx\citep{nutt2014nuttx} and FreeRTOS\citep{freertos2003}, perform well in real-time scheduling.
Some of them also support memory protection to improve the security\citep{wang2017embedded}.
However, none of them has provided a formal correctness proof of its source code.

A potential source of software failures lies in the implementation of device drivers.
The driver has to rely on the behavior of that device, for instance, to tell when it is ready to read or write data, or whether a previous write is complete or not.
However, due to the complexity of modern hardware, it is difficult to consider all possible abnormal situations when implementing the device driver.
For example, it is common for a driver to loop until some status bit on the device is set.
If the device does not update this bit in time, then this delays the execution of the driver, and potentially blocks the whole system if the driver runs in the kernel mode and is not interruptible.


The main contribution of this paper includes a new software architecture for improving the reliability and safety of drone systems at the source code level by introducing formal verification techniques.
In particular, the proposed architecture is based on CertiKOS (Certified Kit Operating System)\citep{gu2015deep}, which enjoys a formal functional correctness guarantee. We adopt this methodology and formally verify the device driver for a drone control system layer by layer, and demonstrate that this indeed improves its safety and reliability.
The same architecture could be extended to autonomous cars, home service robots and other safety-critical systems.

This paper is organized as follows.
Section \textbf{2} describes the proposed software architecture.
Section \textbf{3} describes the formal verification of driver modules.
Experiments and discussions are presented in Section \textbf{4}.

\section{HIERARCHICAL SOFTWARE ARCHITECTURE}
\begin{figure}[htb]
\centering
\includegraphics[height=6.5cm]{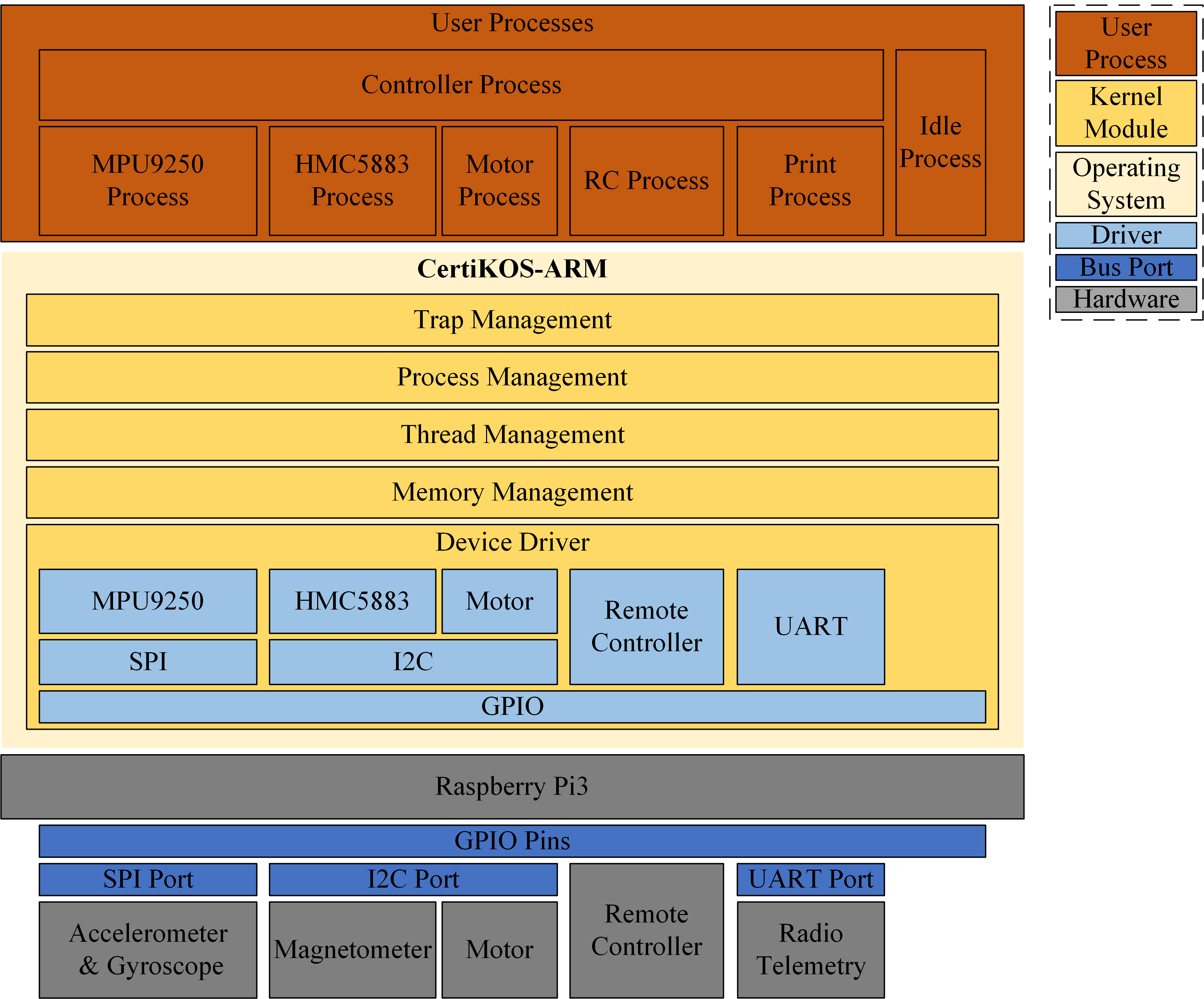}
\caption{Overall software architecture}
\label{arch}
\end{figure}

In order to improve the reliability and safety of the software stack of a drone system, a new hierarchical software architecture is proposed (as shown in Fig. \ref{arch}).
In this architecture, an operating system kernel, CertiKOS\citep{chen2016toward}, plays the central role of managing devices such as motors and sensors, and scheduling user tasks such as the control loop, the sensor fusion program, etc.

A Raspberry Pi3 board is equipped on the drone as its main controller, which connects with multiple sensors and actuators through general purpose I/O (GPIO) pins.
CertiKOS-ARM, the ARM port of CertiKOS, is installed on the board to manage these devices, either directly or through bus drivers, and to expose them to user space programs.
During each control period, the sensor fusion algorithm reads from sensors to generate a reliable attitude estimation.
Then the controller decides its next movement and writes control signals to corresponding motors.
There is also an RC task which reads the receiver to get control signals from the remote controller.
In this way, the reliability of a drone system depends heavily on the correctness of its device drivers.

In the proposed architecture, all software modules including the kernel and device drivers should be formally verified in order to ensure the functional correctness of their source code.
CertiKOS has been formally verified on x86 in previous work \citep{gu2015deep}, and its implementation has been ported to the ARM architecture successfully.
This paper mainly focuses on verification of device drivers for the drone system. It relies on the partially verified CertiKOS-ARM, which includes modules for memory management (verified), thread management (not verified), etc.

\section{DRIVER VERIFICATION}
In a typical drone control system, it is necessary and important to estimate the drone attitude accurately.
Raw data for the drone attitude estimation are usually provided by three sensors: accelerometer, gyroscope and magnetometer.
In our system, the accelerometer and gyroscope depend on the Serial Peripheral Interface (SPI) bus to transmit sensing signals, and the magnetometer uses the Inter-Integrated Circuit (I2C) bus.

\begin{figure}[!htb]
\small
\centering
\includegraphics[scale=0.35]{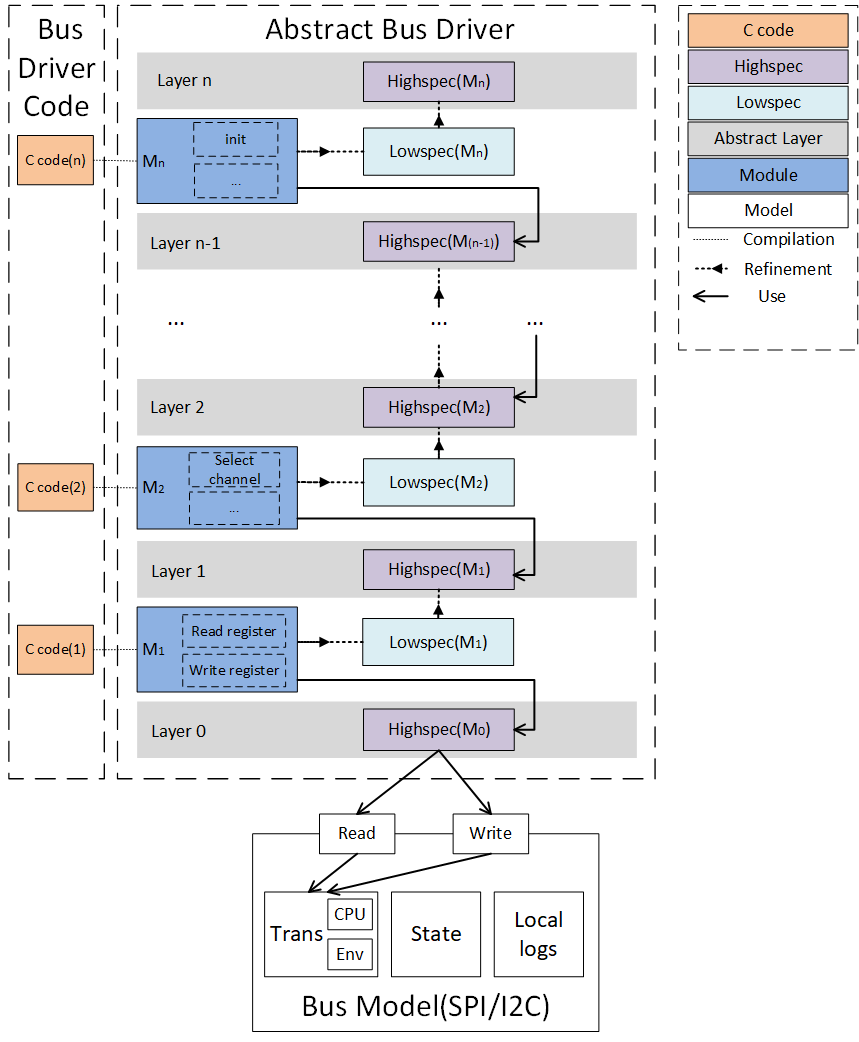}\\
\caption{Driver verification structure}
\label{veri_stru}
\end{figure}

Following the same methodology as presented in \citep{chen2016toward}, the driver verification can be divided into three phases.
Firstly, we build a bus model which abstracts machine registers and the physical memory into a state transition system.
Afterwards, we define an abstract interface for reading and writing the bus, as shown in Fig. \ref{veri_stru}.

During the second phase, we divide the C code of the device driver into multiple layers according to their functionalities and dependencies, as shown in Fig. \ref{veri_stru}.
We further convert these individual C functions into their corresponding Clight\citep{leroy2009formal} abstract syntax tree, so that we can reason about their behaviors by utilizing the Clight semantics (It is actually an extended semantics as detailed in \citep{gu2015deep}).
The set of abstract syntax tree implementing a layer is called a module, i.e. $M_{n}$ in Fig. \ref{veri_stru}.

Next, we abstract each C function into a Coq function (which is called a specification or a primitive), while still capturing everything we want to know about the behavior of its source code.
This is achieved by following the approach of deep specifications \citep{gu2015deep}.
Then we define invariants for each layer, and prove that all primitives preserve these invariants , so that the higher layer will only operate on valid states of its underlay.

It is not straightforward to prove the refinement between the module and its specification (Highspec) in one step.
Hence, we follow \citep{gu2015deep} and introduce the Lowspec to bridge the gap.
While Highspec focuses on the abstract states and high-level invariants, the Lowspec deals with the memory state and low-level invariants.
The set of Highspecs constitutes the abstract layer of the corresponding module, which is relied upon by other modules.
On the other hand, Lowspec is only used for simplifying the refinement proof and hidden from higher layers.

The final phase is the verification of each driver, based on the bus model and abstract bus driver layers obtained from the first two steps. Two refinements have to be proved in each abstract layer\citep{gu2015deep}:
<1> the refinement from Lowspec to Highspec;
<2> each module correctly implements its Lowspec.
If any of the conditions are not satisfied, we adjust either the Lowspec or the original C code until all modules are verified.
Then the deep specification framework links the verification of all layers together to achieve a verified program.
This guarantees the functional correctness of our adjusted C code of device drivers, which in turn contributes to the reliability and safety of the drone system.

\subsection{The bus model}

The characteristics of the bus in an actual physical system depend on the I/O operations of the CPU and its interactions with the external sensor.
Hence, the SPI/I2C bus could be modeled as finite state transition systems interacting with the CPU and external sensors.
Different I/O operations or external sensor events lead to different corresponding changes to the state.
Bus transitions (i.e. \texttt{Trans} in  Fig. \ref{veri_stru}) therefore include these two types of interactions \citep{chen2016toward}.

Notice that the CPU carries out read/write operations on bus registers through the I/O command. We model these operations on both the SPI and I2C bus as in Definition \ref{cpu_model}.

\begin{definition}[CPU Operation on Bus]
\label{cpu_model}
\[
\begin{array}{l}
\mathcal{O} \verb+::= input n+\\
\verb+      | output n v+\\
\end{array}
\]
\end{definition}

$\mathcal{O}\verb+::= input n+$ denotes reading a value from the register whose address is \texttt{n}. $\mathcal{O}\verb+::= output n v+$ means the CPU writes a value \texttt{v} to the register at address \texttt{n}.

The following subsections describe definitions of the state machine for I2C and SPI, and how they are updated by CPU operations and external actions.

\subsubsection{The I2C bus model}

In order to formally define the I2C bus model, we first construct its abstract state.

\begin{definition}[The I2C bus abstract state]
\label{i2c_state}
$$
\begin{array}{l}
\verb+Record I2CState  :=+\\
\verb+  mkI2CState {+\\
\verb+            I2C_OA: +\mathbb{Z}\\
\verb+            I2C_SA: +\mathbb{Z}\\
\verb+            I2C_RX_DATA: +\mathbb{Z}\\
\verb+            I2C_TX_DATA: +\mathbb{Z}\\
\verb|            ...|\\
\verb+            }+\\
\end{array}
$$
\end{definition}

Although the physical bus hardware is sophisticated and contains many more states and operating modes, most of them are irrelevant regarding the attached sensor, such as the 10-bits addressing mode, high-speed mode, etc. Therefore, we fix its operation to the 7-bit addressing mode, and only abstract 10 registers (Definition \ref{i2c_state}) to formalize the state of a physical I2C bus. These include the base address of the device interface state \texttt{I2C\_OA} and slave address state \texttt{I2C\_SA}, which serve as identities when connecting to specific devices. We also model the data receiving buffer state \texttt{I2C\_RX\_DATA} and the data sending buffer state \texttt{I2C\_TX\_DATA} to describe the read/write buffer in a real I2C bus, etc.

Based on Definition \ref{cpu_model} and Definition \ref{i2c_state}, we define the CPU's read/write operations for I2C bus as Definition \ref{i2c_state_transition}.

\begin{definition}[I2C state transition function based on CPU operation]
\label{i2c_state_transition}
\small
$$
\begin{array}{l}
\delta^{\texttt{CPU}}_{\texttt{I2C}} \verb+(op: +\mathcal{O} \verb+) (s: I2CState) : I2CState := +\\
\verb+    |   op = input n -> s+\\
\verb+    |   op = output n v ->+\verb+ s{I2C_OA: v}, if n = I2C_OA+\\
\verb+                                   s{I2C_SA: v}, if n = I2C_SA+\\
\verb+                                   ...+\\
\end{array}
$$
\end{definition}

Function ${\delta^{\texttt{CPU}}_{\texttt{I2C}}}$ describes the interaction between CPU and I2C bus, which takes the CPU operation $\verb+op: +\mathcal{O}$ and the current state as arguments, and returns the resulting state after this operation. A read operation ($\verb+op= input n+$) does not change the I2C state. A write operation ($\verb+op= output n v+$) updates the corresponding field in the abstract state to \texttt{v}.

As mentioned previously, besides I/O operations issued by the CPU, external sensor events may also affect the state of I2C bus. There are three kinds of events for I2C bus as listed in Definition \ref{i2c_event}: non-event, acknowledgment responding event and data receiving event.

\begin{definition}[I2C external sensor event]
\label{i2c_event}
\small
$$
\begin{array}{l}
\texttt{E}^{\texttt{env}}_{\texttt{I2C}} \verb+::= NullEvent+\\
\verb+          | ACKEvent+\\
\verb+          | RecvEvent(val: + \mathbb{Z} \verb+)+\\
\end{array}
$$
\end{definition}

\texttt{NullEvent} represents a non-event in which the I2C bus is waiting for other functional events. \texttt{ACKEvent} represents the acknowledgment responding event in which the I2C bus receives an acknowledgment. \texttt{RecvEvent} denotes the data receiving event in which the I2C bus receives an integer data \texttt{val}.

Based on Definition \ref{i2c_state} and Definition \ref{i2c_event}, we model state transitions of the I2C bus triggered by external events as Definition \ref{i2c_tran_ext}.

\begin{definition}[The I2C state transition function based on external sensor events]
\label{i2c_tran_ext}
\small
$$
\begin{array}{l}
\delta^{\texttt{env}}_{\texttt{I2C}}\verb+(e: + E^{\texttt{env}}_{\texttt{I2C}}\verb+)(s: I2CState) : I2CState :=+\\
\verb+          | s, if e = NullEvent+\\
\verb+          | s, if e = ACKEvent+\\
\verb+          | +\texttt{s}'\verb+, if e = RecvEvent(val)+\\
\end{array}
$$
\end{definition}

The acknowledgment responding event and the non-event will not change the I2C state. For receiving event, the I2C bus receives an integer data \texttt{val}, and copies this value to the register \texttt{I2C\_RX\_DATA} as shown in the following function:

\begin{small}
$$
\begin{array}{l}
\delta^{\texttt{env}}_{\texttt{I2C}}\verb+(e: + E^{\texttt{env}}_{\texttt{I2C}}\verb+)(s: I2CState) : I2CState :=+\\
\verb+        | s, if e = NullEvent+\\
\verb+        | s, if e = ACKEvent+\\
\verb+        | s{I2C_RX_DATA: val}, if e = RecvEvent(val)+\\
\end{array}
$$
\end{small}

Notice that in the I2C bus model, an external sensor event list $l^{\texttt{env}}_{\texttt{I2C}}$ is also constructed to decide the order of all events being processed by the CPU. At the same time, a local event log, Fig. \ref{veri_stru}, is set up to record events which are already processed in the event list.

Once state transition functions of the I2C bus model are defined, we connect transitions caused by CPU operations with transitions triggered by external events to model the overall effect of reading/writing the I2C bus. And they constitute the interface for the device driver to interact with the I2C bus.


\begin{definition}[The I2C bus read semantics]
\label{i2c_read}
\small
$$
\begin{array}{l}
\verb+(e,+ l'_{i} \verb+) = next(+ l^{\texttt{env}}_{\texttt{I2C}} \verb+,+ l_{i} \verb+)+\\
\texttt{s}' \verb+ = + \delta^{\texttt{env}}_{\texttt{I2C}} \verb+(s,e)+\\
\verb+res = + \kappa \verb+(n,+ \texttt{s}'\verb+)+\\
\texttt{s}''\verb| = | \delta^{\texttt{CPU}}_{\texttt{I2C}} \verb|(| \texttt{s}' \verb|,(input n))|\\
\end{array}
$$
\end{definition}

In Definition \ref{i2c_read}, we first find the next event \texttt{e} to handle by comparing the event list $l^{\texttt{env}}_{\texttt{I2C}}$ with local event log $l_{i}$, which is denoted by the function $\texttt{next(}l^{\texttt{env}}_{\texttt{I2C}}\texttt{,}l_{i}\texttt{)}$.
Then, we apply the I2C state transition function $\delta^{\texttt{env}}_{\texttt{I2C}}$ on the event \texttt{e} and the current I2C state \texttt{s} to obtain the next I2C state $\texttt{s}'$.
The next step is to obtain the value \texttt{res} from the abstract state $\texttt{s}'$ and register address \texttt{n}.
Finally, we update the I2C state again through the state transition function $\delta^{\texttt{CPU}}_{\texttt{I2C}}$.
Given all above premises, semantics of reading the I2C bus is defined as: $\texttt{read(n,s,}l_{i}\texttt{,}l^{\texttt{env}}_{\texttt{I2C}}\texttt{)=(res, }\texttt{s}''\texttt{, }l'_{i}\texttt{)}$.

Similarly, the following defines the write semantics on the I2C bus.

\begin{definition}[The I2C bus write semantics]
\label{i2c_write}
$$
\inferrule
{
	{
		{
			\begin{array}{l}
			\texttt{(e},l'_{i}\texttt{) = next(}l^{\texttt{env}}_{\texttt{I2C}}\texttt{,}l_{i}\texttt{)}\\
			\texttt{s}'\texttt{ = }{\delta^{\texttt{env}}_{\texttt{I2C}}}\texttt{(s,e)}\\
			\texttt{s}''\texttt{ = }{\delta^{\texttt{CPU}}_{\texttt{I2C}}}\texttt{(}\texttt{s}'\texttt{,(output n v))}
			\end{array}
		}
	}
}
{
	\texttt{write(n,v,}l_{i}\texttt{,}l^{\texttt{env}}_{\texttt{I2C}}\texttt{)=(}\texttt{s}''\texttt{,}l'_{i}\texttt{)}
}
$$
\end{definition}

This concludes the definition of the I2C bus model, which is relied upon by the verification of device drivers explained in the next subsection.

\subsubsection{The SPI bus model}

The SPI bus is modeled by the same approach.


\begin{definition}[The SPI bus abstract state]
\label{spi_state}
\small
$$
\begin{array}{l}
\verb+Record SPIState  :=+\\
\verb+  mkSPIState {+\\
\verb+            SpiRx: +\mathbb{Z}\\
\verb+            SpiTx: +\mathbb{Z}\\
\verb+            SpiEn: +bool\\
\verb+            SpiMs: +\texttt{SPI\_MS}\\
\verb|            ...|\\
\verb+            }+\\
\end{array}
$$
\end{definition}

In the SPI Bus model, integer elements \texttt{SpiRx} and \texttt{SpiTx} represent the data receive buffer and data transmit buffer of the actual physical SPI bus, which are abstracted from the data receive register and the transmit register, respectively. The boolean field \texttt{SpiEn} is an abstraction for modeling SPI enabling status. In summary, the SPI bus abstract state contains a total of 25 fields, and they are used in our drone control system.

\subsection{Layer structure of the driver code}

As mentioned at the beginning of this section, we follow \citep{chen2016toward} and divide the bus driver code into layers based to their functionalities and dependencies to enable the compositional verification. Three principles are followed during this process:
<1> similar functions, such as read/write a register, should be put in the same layer;
<2> one layer should not contain too many functions, to make the proof easier;
<3> such layering should not change the overall behavior of the source code.
We show the layer structure of the SPI bus driver, while the layering of the I2C bus driver is similar.

\begin{figure}[htb]
\centering
\includegraphics[height=6cm]{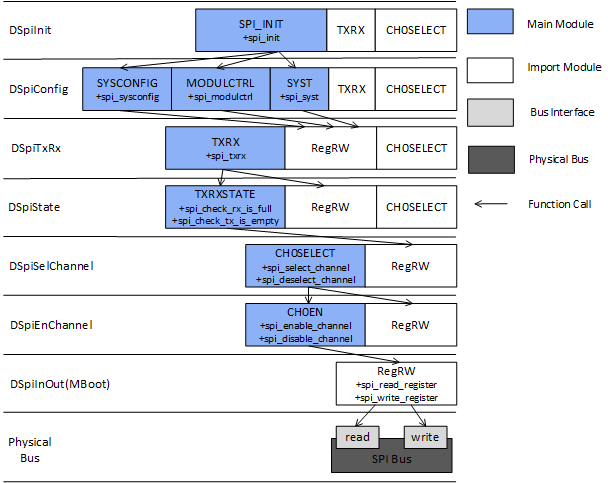}
\caption{Layering structure of SPI bus driver}
\label{spi_layer}
\end{figure}

In Fig. \ref{spi_layer}, each block represents one module in the layer.
For example, in the layer \texttt{DSpiInOut}, the module \texttt{RegRW} contains two functions for reading and writing registers.
The arrow between two modules indicates the calling relation between them, and a module is only allowed to call modules in the lower layer.
For example, the module \texttt{CH0EN} points to (invokes) the module \texttt{RegRW}, and the module \texttt{RegRW} points to the read and write interface of the SPI bus.
The blue block indicates that functions in this module depend on at least one function in another module.
The white block represents the module which is passed through from a lower layer without any modification.
For example, the module \texttt{RegRW} in \texttt{DSpiEnChannel} is passed through directly from the layer \texttt{DSpiInOut}.
The module \texttt{RegRW}, which consists of the read and write interface of the SPI bus, is located at the bottom of the layer architecture.

We discuss the verification of these modules in the next section.

\subsection{Verification of the driver}

In this subsection, we follow the methodology proposed in \citep{gu2015deep} to verify the SPI driver.

\subsubsection{Functional correctness of the C code}

We show the C code for enabling the channel and its corresponding Clight representation in Fig. \ref{example}. The main operation of the function is to write value \texttt{ENABLE\_CHANNEL} to the address \texttt{CH0CTRL} in order to enable the SPI bus.


\begin{figure}[htb]
\begin{small}
$$
\begin{array}{l}
\verb+void mcspi_enable_channel (void) { +\\
\verb+write_register(ENABLE_CHANNEL, CH0CTRL);+\\
\verb+}+\\
\verb+Definition mcspi_enable_channel :=+\\
\verb+(Scall None+\\
\verb+(Evar MCSPI_write_register (Tfunction+ \\
\verb+(Tcons tuint (Tcons tuint Tnil)) tvoid cc_default))+\\
\verb+((Econst_int (ENABLE_CHANNEL) tint) ::+\\
\verb+(Econst_int (CH0CTRL) tint) :: nil))).+\\
\end{array}
$$
\end{small}
\caption{The C source code and its Clight representation (in Coq) of function mcspi\_enable\_channel}
\label{example}
\end{figure}

\begin{figure}[htb]
\centering
\includegraphics[height=7cm]{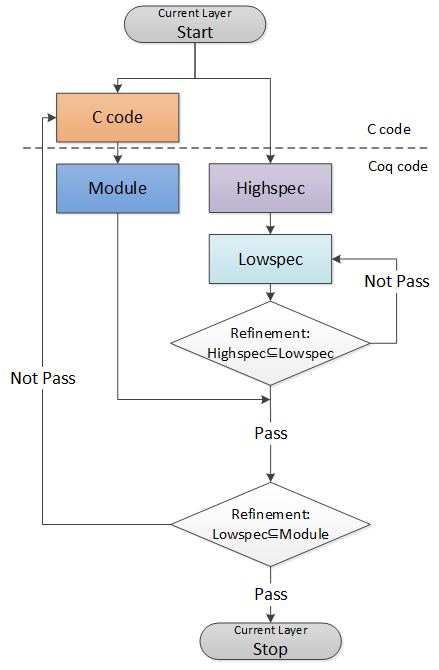}
\caption{Contextual refinement verification for a C function}
\label{modveri}
\end{figure}

The workflow of proving the functional correctness of a module is elaborated in Fig. \ref{modveri}. Clightgen, provided by Compcert\citep{leroy2009formal}, is used to translate the C code of SPI driver into a Clight abstract syntax tree. Then we write the Highspec and Lowspec of the corresponding module in Coq (see Fig. \ref{modveri}) to establish the refinement relation.

The Highspec describes the desired functionality of this module.
For example, the above function \texttt{mcspi\_enable\_channel} is abstracted as below.

\begin{small}
$$
\begin{array}{l}
\verb+Function + \hat{\sigma}_{\texttt{mcspi\_enable\_channel}} \verb+(abs: RData)+\\
\verb+                        : option RData :=+\\
\verb+  match(spi abs)with+\\
\verb+  | SpiState _ SpiEN.en _ -> +\\
\verb+         Some(abs{spi:(SpiState _ SpiEn.Enable _)})+\\
\verb+  | _ -> None+\\
\verb+  end+
\end{array}
$$
\end{small}

Here, \texttt{RData} contains all states of the system, such as the page table, the process control block, etc. This function only updates \texttt{spi}, and \texttt{abs.SpiState} is an instance of \texttt{spi}. The enable bit(\texttt{SpiEn}) of the SPI bus state(\texttt{SpiState}) will be changed from the previous value to \texttt{Enable}, which describes the behavior of the SPI enable operation in the original C code.

The Lowspec also abstracts the behavior of each function in this module, but is specified in a way that is closer to the concrete hardware. In the case of enabling the SPI bus, it looks very similar to the corresponding Highspec because only function invocation is involved. The following is the low specification of the function \texttt{mcspi\_enable\_channel} written in Coq:

\begin{small}
$$
\begin{array}{l}
\verb+Inductive + \hat{\sigma}\texttt{LOW}_{\texttt{mcspi\_enable\_channel}} \verb+(abs +\texttt{abs}'\verb+:RData)+\\
\verb+                                           (m0:mem) :=+\\
\verb+|+ \hat{\sigma}_{\texttt{mcspi\_enable\_channel}} \verb+abs = Some + \texttt{abs}' \verb+->+\\
\hat{\sigma}\texttt{LOW}_{\texttt{mcspi\_enable\_channel}}\verb+(m0,abs) -> (m0,+\texttt{abs}' \verb+).+\\
\end{array}
$$
\end{small}

Here, \texttt{RData} represents the abstract state and \texttt{mem} represents the memory state. Notice that the memory state, \texttt{m0}, does not change since this function does not involve any direct memory operation. Thus, the overall behavior is a transition from \verb+(m0,abs)+ to \verb+(m0,abs')+.
Then we prove the refinement between the Highspec and the Lowspec defined as follows:

\begin{figure}[!h]
\label{refinement}
$$
\begin{array}{l}
\text{Highspec} \subseteq \text{Lowspec} :=\\
\forall a,a',m, ~(a \xrightarrow[]{\text{Highspec}}a') \wedge (a \sim m)\\
\Rightarrow \exists m', ~(m \xrightarrow[]{\text{Lowspec}} m') \wedge (a' \sim m')
\end{array}
$$
\end{figure}

The Highspec and Lowspec may operate on different types of states, so that we use $a$, $a'$ and $m$, $m'$ to distinguish between the two. However, we establish a relation $a\sim m$ between two states on these two different levels, which holds only if $a$ is a valid abstraction of $m$. This refinement relation states that if the Highspec takes one step from $a$ to $a'$, and that its initial state $a$ is a proper abstraction of $m$, then the corresponding Lowspec must be able to step from $m$ to $m'$, where the $\sim$ relation also holds between $a'$ and $m'$.

Similarly, we prove the refinement relation between the Lowspec and the actual C code. Combined together, we get the refinement from the Highspec to the actual C code, which is exactly its functional correctness proof.

As shown in Fig. \ref{modveri}, it is possible during the verification process that we find certain refinement relations do not hold. This is either due to a flaw in the specification, which we need to revise and try again, or is indeed caused by a bug in the source code. In the latter case, we have to fix the bug so that the functional correctness of the source code could be verified.

\subsubsection{Linking all layers together}

\begin{figure}[!htb]
	\centering
	\includegraphics[height=4cm]{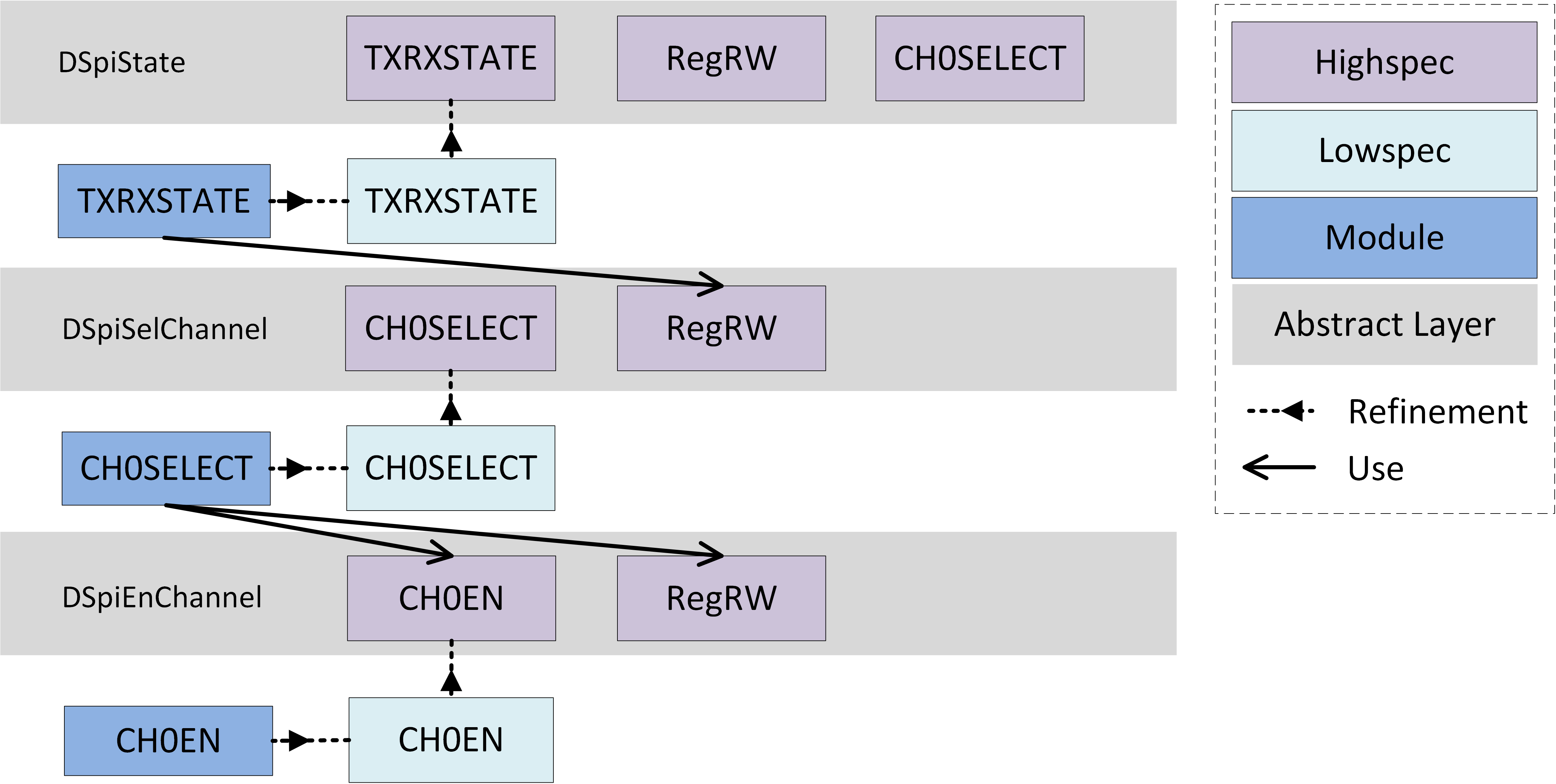}
	\caption{Verification of the SPI driver}
	\label{layerveri}
\end{figure}

The functional correctness proof of each layer assumes the functional correctness of the layer below it. Part of the layer architecture of the SPI bus driver is presented in Fig. \ref{layerveri} to illustrate how we build up the verification layer by layer.

The module \texttt{CH0EN} is first verified, meaning the behavior of its C code indeed follows its specification. It then serves as the interface of layer \texttt{DSpiEnChannel}, which is invoked by layer \texttt{DSpiSelChannel}. Similarly, the layer \texttt{DSpiSelChannel} also exposes the Highspec \texttt{CH0SELECT} as part of its interface, which could be used by upper layers.

\begin{figure}[!htb]
	\centering
	\includegraphics[scale=0.42]{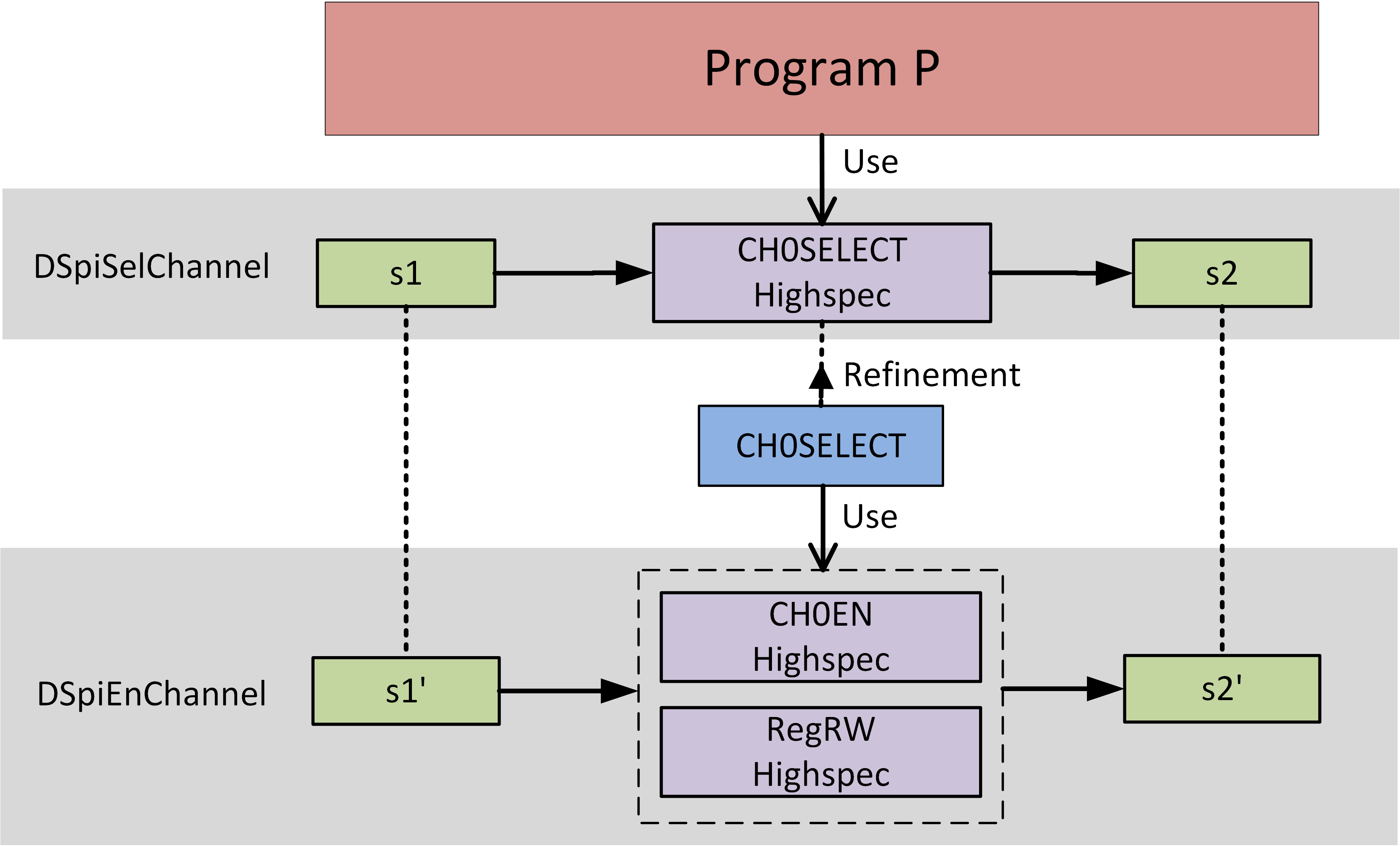}
	\caption{Refinement between abstract layers}
	\label{sim}
\end{figure}

The framework we use \citep{gu2015deep} enables us to link layers together and prove the following contextual refinement between layers. Assume that \texttt{P} is a program which uses the function \texttt{CH0SELECT}.
As in Fig. \ref{sim}, the behavior of linking \texttt{P} with the module \texttt{CH0SELECT} (written as $\texttt{P} \oplus \texttt{CH0SELECT}$) and running them on the layer \texttt{DSpiEnChannel} is equivalent to the behavior of running program \texttt{P} on the layer \texttt{DSpiSelChannel} (written as \texttt{P$@$DSpiSelChannel}). We can write the refinement between these two executions as follows:

\begin{center}
$\texttt{P}@\texttt{DSpiSelChannel} \subseteq \texttt{P} \oplus \texttt{CH0SELECT}@\texttt{DSpiEnChannel}$.
\end{center}

Once this refinement is proved, the actual implementation of the function \texttt{CH0SELECT} is hidden under the layer \texttt{DSpiSelChannel}, while we are still able to reason about all behaviors of program \texttt{P}.

Finally, we use Compcert \citep{leroy2009formal} to generate assembly code for all verified modules, which carries the functional correctness property all the way down to the assembly code level \citep{gu2015deep}.

\section{EXPERIMENTS}
\subsection{Methods and procedures}

\begin{figure}[!htb]
	\centering
	\includegraphics[height=5cm]{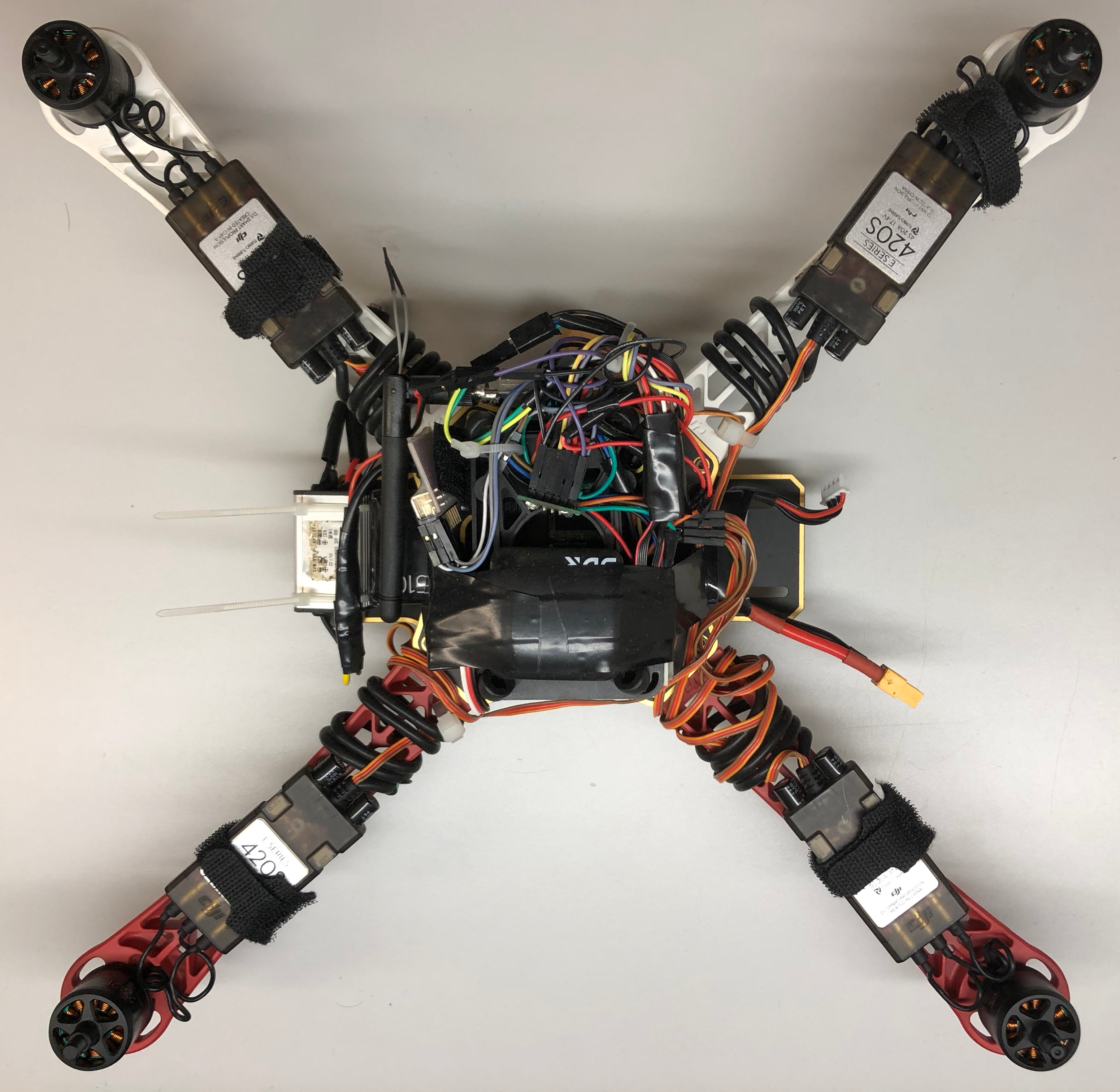}
	\caption{Drone used in experiments}
	\label{drone_platform}
\end{figure}

A drone (Fig. \ref{drone_platform}) was built for all experiments. Three basic
sensors, including an accelerometer, a gyroscope and a magnetometer were used to
estimate the attitude of the drone. Their configurations are listed in Table
\ref{Table:sensordata}. A radio telemetry was used to record the flight data.
Experiments were designed in this section to simulate erroneous situations or
bugs of bus drivers. We set up the system so that bugs occur every 5-10 seconds,
whose effect is to delay the execution of the driver code for as long as 0.2
seconds.
This simulates the situation when the driver keeps polling for new data
without enforcing any timeout mechanism.
In this case, an anomaly in the device may block the driver for a long time,
which in turn blocks the execution of the whole system.

\begin{table*}[thp]\footnotesize
	\centering
	\caption{Configurations of three sensors of the drone$^\textbf{*}$} \label{Table:sensordata}
	\addtolength{\tabcolsep}{4.8pt}
	\begin{tabular*}{14cm}{lcccc}
		\toprule[0.75pt]
		Sensor				&Chip Name		&Measurement Range		&Sensitivity		&Sampling Rate\\
		\midrule[0.5pt]
		Accelerometer		&MPU9250		&$\pm$8 g				&4096 LSB/g			&200 Hz\\
		Gyroscope			&MPU9250		&$\pm$1000 dps			&32.8 LSB/dps		&200 Hz\\
		Magnetometer		&HMC5883		&$\pm$1.3 Gs			&1090 LSB/Gs		&75 Hz\\
		\bottomrule[0.75pt]
		\multicolumn{5}{p{12cm}}{\scriptsize $^*$ Taken from datasheets of MPU9250 and HMC5883.

			g: standard gravity; dps: degree per second; Gs: gauss; LSB: least significant bit}
	\end{tabular*}
\end{table*}

Two drone systems are tested in the real field and the results are further compared.
The first system is the drone system with a verified SPI bus driver as explained in the previous section.
The second one is a system with unverified SPI bus driver.
Both of these two systems are equipped with the verified I2C bus driver.

Ten trials have been carried out with different bugs randomly occurring in the SPI bus driver.
We record and compare attitudes of the drone (roll, pitch and yaw) since they are the most critical metrics to its safety.
The attitudes are computed by the same gradient descent method\citep{madgwick2011estimation} using IMU data read from the SPI bus.

\subsection{Results and discussions}

\begin{figure}[!tb]
	\centering
	\includegraphics[height=6.9cm]{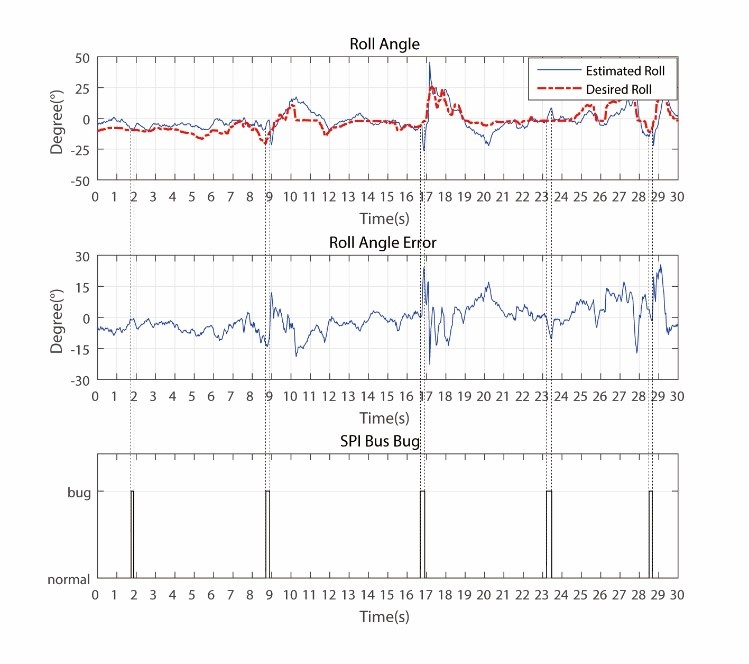}
	\caption{Roll angle response of drone with unverified SPI bus driver}
	\label{rolldata}
\end{figure}
\begin{figure}[!tb]
	\centering
	\includegraphics[height=6.9cm]{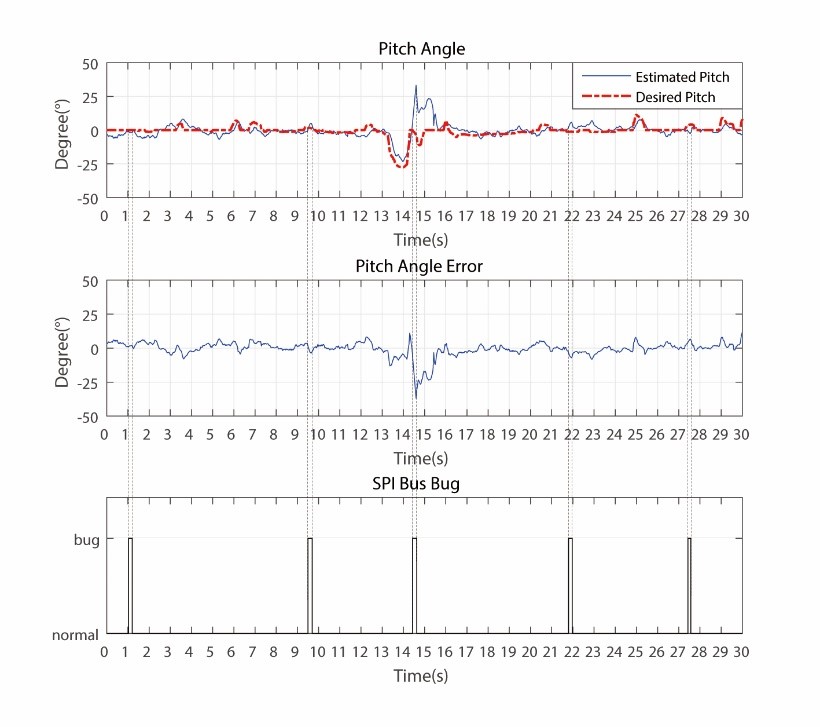}
	\caption{Pitch angle response of drone with unverified SPI bus driver}
	\label{pitchdata}
\end{figure}
\begin{figure}[!tb]
	\centering
	\includegraphics[height=6.9cm]{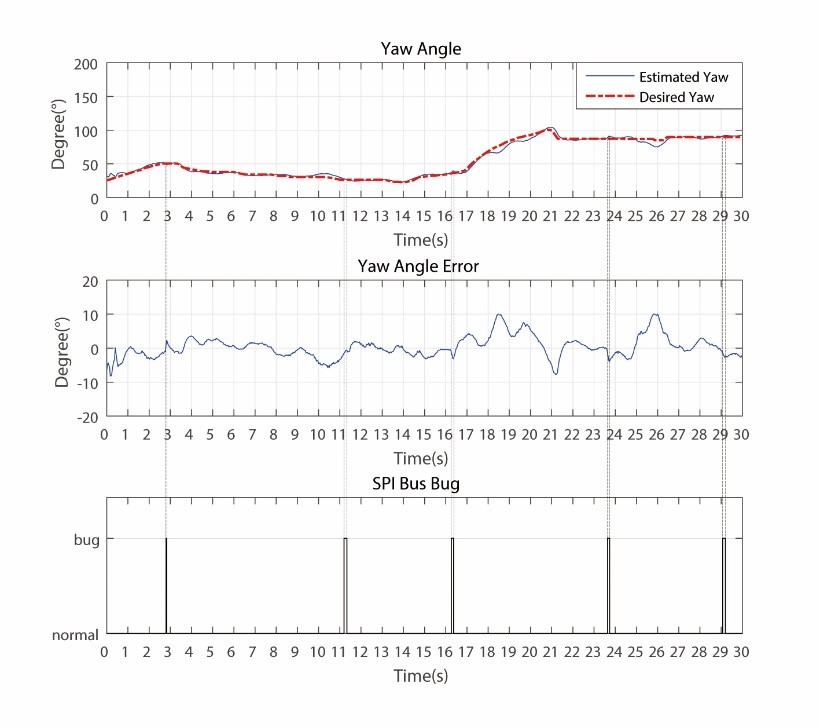}
	\caption{Yaw angle response of drone with unverified SPI bus driver}
	\label{yawdata}
\end{figure}
\begin{figure}[!tb]
	\centering
	\subfigure[]{\includegraphics[height=3.2cm]{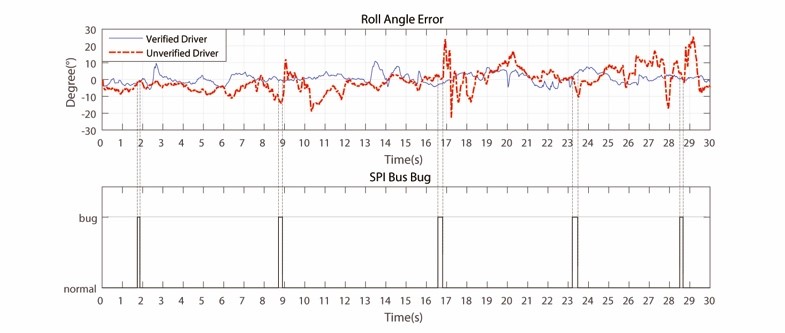}}
	\subfigure[]{\includegraphics[height=3.2cm]{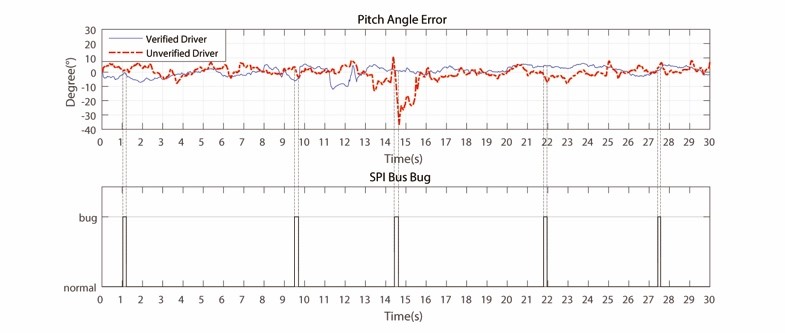}}
	\caption{Comparison of attitude errors in two drone systems}
	\label{comp}
\end{figure}

Fig.~\ref{rolldata} shows the roll angles of the unverified drone system. Solid
lines in subfigure `\texttt{Roll Angle}' represent the computed actual roll angles while
dashed lines represent the desired values required by the remote controller. The
differences between the actual and desired value (errors) are shown in subfigure `\texttt{Roll
Angle Error}'. Three peaks of errors are observed in the timeline 8.6s, 16.8s and
28.5s. At these time intervals, software bugs in the device drivers cause
delayed process and response of sensor data, which further blocks the controller's execution
for the next multiple control periods. Software bugs are also detected at
1.8s and 23.2s in the timeline (subfigure `\texttt{SPI Bus Bug}'). However, these bugs have no
obvious impact on the roll angle, due to the relatively steady attitude of the drone.
When these bugs occur, the input of each motor will be the same as it in the previous period.
If the current attitude of the drone doesn't change a lot comparing with the previous one, the drone will stay stable by using the same motor input.
The same phenomenon exists on the pitch angle as shown in Fig.~\ref{pitchdata}.

Fig.~\ref{yawdata} shows the value of the yaw angle, which does not experience the same variation upon software faults caused by bugs.
It is attributed to the sensor fusion algorithm, which uses data from both the IMU
(connected with the SPI bus) and the magnetometer (connected to the I2C bus) to
improve the accuracy of estimated yaw angle.

Fig.~\ref{comp} shows comparison of attitude errors between these two drone systems.
The exists of software bugs leads to significant differences between desired and actual
pitch and roll angle.

\begin{figure*}[t]
	\centering
	\subfigure[Snapshots of drone flights]{\includegraphics[width=15cm]{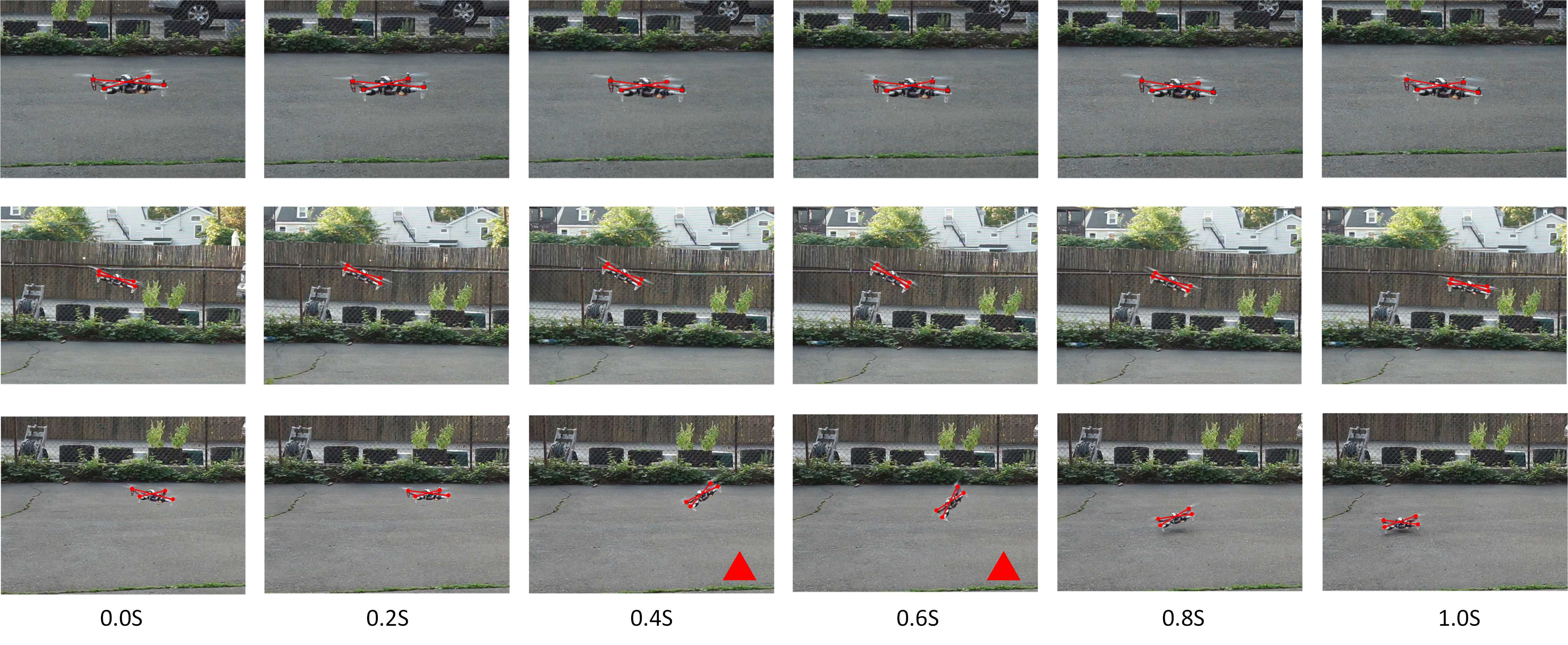}}
	\subfigure[Variation of drone attitude]{\includegraphics[height=2.5cm]{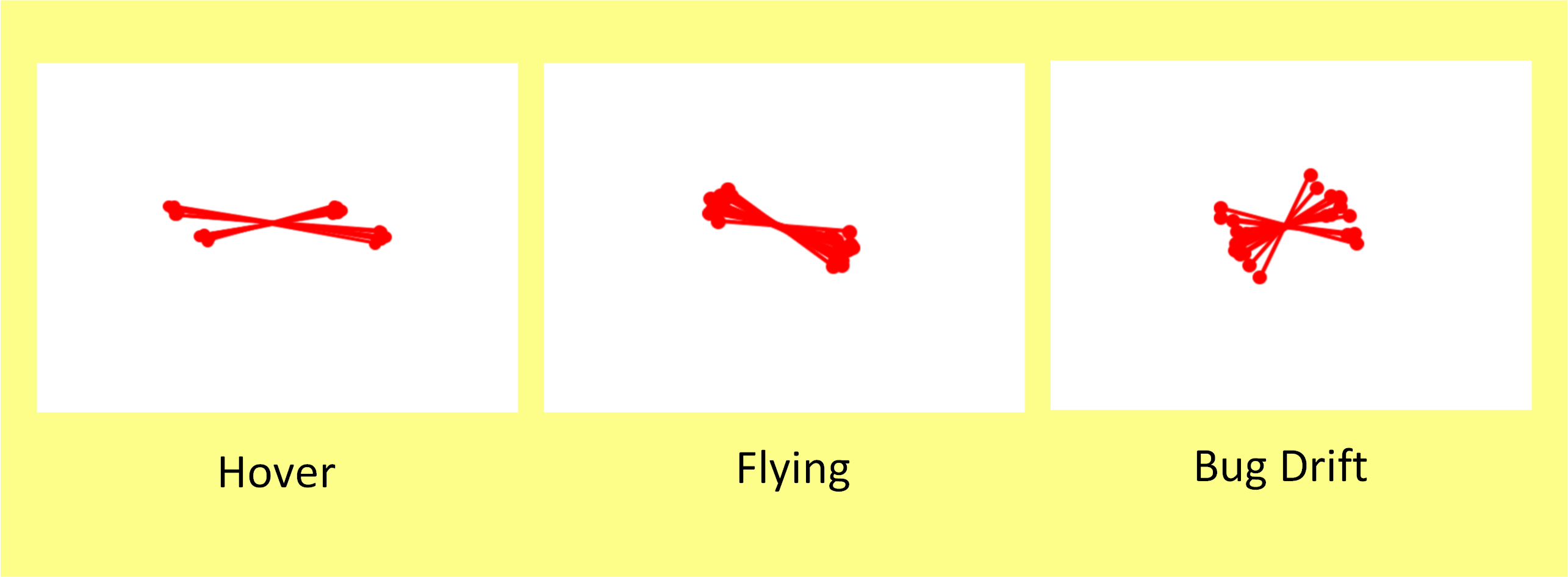}}
	\caption{An empirical comparison between systems with/without a verified SPI bus driver}
	\label{snap}
\end{figure*}

Fig. \ref{snap} shows a series of snapshots for different drone flights in a consequent timeline.
Drones in the first two rows have installed verified device drivers.
It could hover, and is able to change its attitude and fly forward.
The third row shows the situation when there are bugs in the drone's SPI bus driver.
The pictures show greater variations of the drone's attitude compared to the first and second rows, even if they are operated in the same manner.
This demonstrates that bugs in the SPI bus driver indeed degrades the stability of a drone.

\section{CONCLUSIONS}

A new software architecture and development method targeting at safety and reliability for a drone system is proposed in this paper.
With the help of formal verification, several bus drivers which play critical roles in the flight control are formally verified.
Experiments in the filed tests show that the proposed system enjoys the improved reliability by eliminating the subtle bugs that could be introduced in the software development.

In our future work, we plan to extend the proposed architecture with virtualization support.
A hypervisor could be introduced to support third-part systems without compromising the inherited safety and security by enforcing strong isolation and non-interference properties.

\bibliographystyle{fitee}
\bibliography{ref}

\end{document}